\documentclass[bibnotes,twocolumn,showpacs,amsmath,amssymb,nobalancelastpage,dvips]{revtex4}
\usepackage{bm}
\usepackage{graphics,psfig,epsfig,subfigure,afterpage}
\usepackage{amsfonts}
\usepackage{psfrag}

\newcommand{\be}{\begin{equation}}
\newcommand{\ee}{\end{equation}}
\newcommand{\bea}{\begin{eqnarray}}
\newcommand{\eea}{\end{eqnarray}}
\newcommand{\gdot}{\dot{\gamma}}
\newcommand{\gdotbar}{\bar{\dot{\gamma}}}

\newcommand{\eg}{{\it e.g.\/}}

\newcommand{\versus}{{\it vs.\/}}

\newcommand{\bw}{\begin{widetext}}
\newcommand{\ew}{\end{widetext}}

\begin{document}

\title{Spatio-temporal oscillations and rheochaos in a simple model of shear banding}
\author{S. M. Fielding}
\email{s.m.fielding@leeds.ac.uk}
\author{P. D. Olmsted}
\affiliation{Polymer IRC and
  Department of Physics \& Astronomy, University of Leeds, Leeds LS2
  9JT, United Kingdom} 
\date{\today} 
\begin{abstract}
  We study a simple model of shear banding in which the flow-induced
  phase is destabilised by coupling between flow and microstructure
  (wormlike micellar length).  By varying the strength of the
  instability and the applied shear rate, we find a rich variety of
  oscillatory and rheochaotic shear banded flows. At low shear and
  weak instability, the induced phase pulsates in width next to one
  wall of the flow cell.  For stronger instability, single or multiple
  high shear pulses ricochet across the cell. At high shear rates we
  observe oscillating bands on either side of a defect. In some cases,
  multiple such defects exist and propagate across the cell to
  interact with each other. We discuss our results in the context of
  recent observations of oscillating and fluctuating shear banded
  flows.
\end{abstract}
\maketitle

\begin{figure}[t]
\includegraphics[scale=0.33]{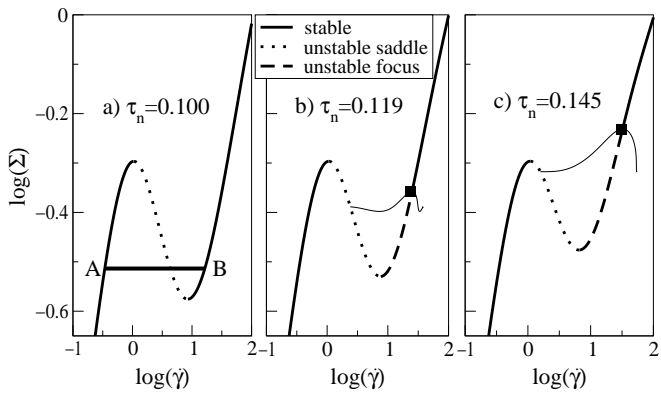}
\caption{Intrinsic constitutive curves for differing degrees of
  coupling between flow and micellar length. a) Weak coupling, giving
  the standard coexistence of stable low and high shear bands (A and
  B); b) moderate coupling; c) strong coupling. Squares show Hopf
  bifurcations.  The thin black lines delimit the periodic orbit of
  the local model at fixed $\Sigma$.
\label{fig:linear} } 
\end{figure}

Complex fluids commonly undergo flow instabilities and flow-induced
phase transitions that result in spatially heterogeneous ``shear
banded'' states.  Classically studied systems include wormlike
micellar surfactants~\cite{BritCall97}; dense lamellar
onion~\cite{diat93} or micellar cubic~\cite{EiserMPD00} phases; and
polymer solutions~\cite{HilVla02}.  Fluidity banding has also been
reported in soft glassy materials \cite{SolLeqHebCat97} such as
colloidal suspensions~\cite{RayMouBauBerGuiCou02} and simulated
Lennard Jones particles~\cite{VarBocBarBer03}.  Experimentally, the
basic observation is of two coexisting shear bands with differing
viscosity and microstructure (or fluidity).  Theoretically, this is
captured by invoking multiple flow branches in the constitutive
relation of shear stress \versus\ shear rate,
$\Sigma(\gdot)$~\cite{olmsted90,spenley93,lu99}.  The system then
separates into a steady state comprising two shear bands, each on its
own flow branch (Fig.~\ref{fig:linear}a).

However an accumulating body of data shows this basic picture to be
oversimplified: Many shear banding systems display {\em oscillations}
or {\em irregular fluctuations suggesting chaos} in their bulk
rheology, rheo-optics or velocimetry. Example systems include onion
phases (at the layering transition~\cite{WunColLenArnRou01} or
showing size oscillations~\cite{salnoteb}); wormlike micelles (WMs)
with a stress plateau in the flow curve~\cite{BanBasSoo00}; shear
thinning WMs showing common stress banding (band normals in the
flow-gradient direction)~\cite{callnote1,callnote2}; shear thickening
WMs showing common stress~\cite{HBP98} or common strain rate
(vorticity)~\cite{FisWheFul02} banding; and polymer
solutions~\cite{HilVla02}.  In contrast, present models predict steady
banded states~\cite{spenley93,olmsted90,lu99}.

In the negligible Reynolds limit relevant to these viscoelastic
materials, the non linearity underlying this erratic response must
arise in the rheological constitutive behaviour of the
system~\cite{CatHeaAjd02}.  Temporal ``rheochaos'' has recently been
studied in homogeneous models of director dynamics in sheared
nematics~\cite{GroKeuCreMaf01}; and of shear thickening systems with a
single-branched flow curve~\cite{CatHeaAjd02}. In many systems,
however, spatial heterogeneity is likely to be a crucial ingredient of
rheochaos.  In this work, motivated by the above experiments, we
introduce the first model of {\em spatio}-temporal rheochaos in shear
banding systems with multi-branched constitutive curves.

Homogeneous flow is unstable in any region of negative constitutive
slope, $d\Sigma/d\gdot<0$.  This is easily seen in models that take
the ``mechanical variables'' ($\Sigma$, $\gdot$) as the relevant
dynamic variables~\cite{Yerushalmi70}.  The system can then separate
into two shear bands, each on its own stable flow branch.  (See
Fig.~\ref{fig:linear}a, for shear thinning gradient-banding systems.)
In more realistic models, the mechanical variables are coupled to
microstructural quantities such as director
orientation~\cite{olmsted90}, polymeric concentration~\cite{FieOlm03}
or micellar length~\cite{catesturner92}. This coupling can destabilize
the rising high shear branch (causing, \textit{e.g.}, tumbling and
wagging in nematics~\cite{doi81}).  Here we construct a simple model
with an unstable high shear branch (Fig.~\ref{fig:linear}b,c) and show
that it has oscillatory and erratic (chaotic) shear banded states at
imposed global shear rate.  Although we have not yet proved the
erratic states to be strictly chaotic, we refer to them as chaotic
hereafter. Our model resembles other globally coupled reaction
diffusion systems, with similar spatio-temporal
behaviour~\cite{nonlin}.  Aradian and Cates are currently studying
spatio-temporal flows of similar models with a single unstable
constitutive branch~\cite{catesnote}.

We use only the minimal ingredients needed to capture the observed
phenomena, so do not address the microscopics of any given system.
However for concreteness we use the language of shear thinning WMs.
There exist numerous reports of apparently unattainable homogeneous
high shear rate branches in such systems, in which the sample flows
erratically or is ejected from the cell (\eg ~\cite{BritCall97}).
This is seldom discussed in detail, but sometimes attributed to
surface instability.  Nonetheless, the possibility of bulk instability
remains.  Indeed, in some WMs the high shear band breaks into
sub-bands~\cite{LerDecBer00}.

For simplicity we consider just one microstructural variable, the mean
micellar length $n$, and define our model by

\vspace{-0.75cm}
\bea
\label{eqn:one}\Sigma&=&\sigma+\eta\gdot,\\ 
\label{eqn:two}\partial_t\sigma&=&-\frac{\sigma}{\tau(n)}+\frac{g[\gdot\tau(n)]}{\tau(n)}+D\partial_y^2\sigma,\\
\label{eqn:three}\partial_t n&=&-\frac{n}{\tau_n}+\frac{N(\gdot\tau_n)}{\tau_n}.
\eea
Eqn.~\ref{eqn:one} gives the uniform shear stress $\Sigma(t)$ in the
non-inertial limit. It comprises a (generally non-uniform)
viscoelastic micellar contribution $\sigma(y,t)$ and a solvent
contribution with viscosity $\eta$. The dynamics of $\sigma$
(Eqn.~\ref{eqn:two}) has a length-dependent relaxation time
\cite{Soltero}
\be
\label{eqn:tauofn}
\tau(n)=\tau_0 \left(\frac{n}{n_0}\right)^\alpha,
\ee
and a steady homogeneous state $\sigma=g[\gdot\tau(n)]$ set by
\be
\label{eqn:g}
g(x)=\frac{x}{1+x^2},
\ee
admitting a constitutive curve of negative slope. The spatial
gradients $D\partial_y^2\sigma$ allow for a smooth interface of width
$l\propto\sqrt{D}$ between the shear bands.

The micellar length $n$ has its own relaxation time $\tau_n$
(Eqn.~\ref{eqn:three}), presumably related to the (unknown) underlying
rates of micellar scission and recombination.  For example, scission
could be enhanced by the agitation of shearing; or recombination aided
by shear-induced alignment of micellar ends~\cite{catesturner92}. We
assume the former, taking
\be
\label{eqn:N}
N(x)=\frac{n_0}{1+x^\beta}.
\ee
Because this decrease in length only becomes important for typical
shear rates $\gdot>1/\tau_n$, the strength of coupling between the
mechanical quantities $(\Sigma,\sigma,\gdot)$ and microstructure $n$
can be tuned by varying $\tau_n$.

Using this model we study flow between two parallel plates at $y=0,L$
with boundary condition $\partial_y\Sigma=0$, using units in which
$n_0=1,\tau_0=1$ and $L=1$.  We set $\alpha=1.2$, $\beta=1.5$, though our
results are qualitatively robust to reasonable variations in these
values.

\begin{figure}[t]
\includegraphics[scale=0.3]{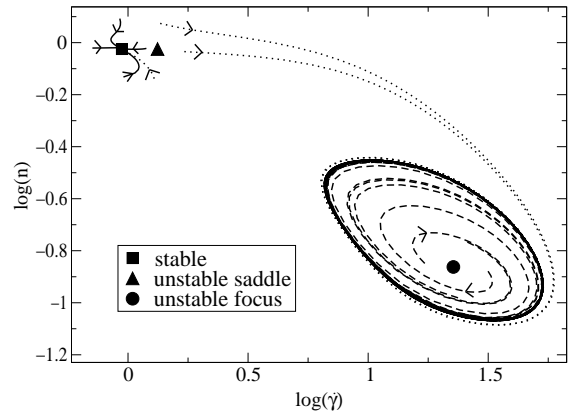}
\caption{Homogenous dynamics ($D=0$)  at fixed $\log\Sigma=-0.301$ for different  initial conditions.  $\tau_n=0.145$.
\label{fig:localFlows} } 
\end{figure}

{\it Homogeneous dynamics---} The constitutive curves $\Sigma(\gdot)$
and domains of instability for homogeneous states are shown in
Fig.~\ref{fig:linear}.  For small $\tau_n\alt 0.115$ we find pure
mechanical instability (Fig.~\ref{fig:linear}a) that does not involve
$n$. The constitutive curve is then (at any fixed $\Sigma$) an unstable
saddle~\cite{Strogatz94} (one unstable eigenvector).  For larger
$\tau_n$, coupling to micellar length broadens this instability into
the rising high shear branch, which is now an unstable
focus~\cite{Strogatz94} (two unstable eigenvectors):
Fig.~\ref{fig:linear}b,c. This instability terminates in a Hopf
bifurcation~\cite{Strogatz94}.

We then studied the non-linear dynamics of the local ($D=0$) model for
fixed $\Sigma$, solving Eqns.~(\ref{eqn:one}-\ref{eqn:three}) via a
fourth order Runge-Kutte method~\cite{PreTeuVetFla92}.  This confirmed
the stability properties of Fig.~\ref{fig:linear}: states near an
unstable (stable) segment of the constitutive curve flow away from
(towards) that segment (Fig.~\ref{fig:localFlows}). They also reveal a
periodic orbit about the unstable high shear branch for stresses just
below the Hopf bifurcation.  We also used AUTO~\cite{AUTO} to trace
the amplitude of the periodic orbits, Fig.~\ref{fig:linear}b,c.
Periodic orbits are the most complex behaviour possible for the local
model since it has only two degrees of freedom, $d=2$.  Chaos requires
$d\ge 3$~\cite{Strogatz94}.

\begin{figure}[tbp]
  \centering 
\hspace{-0.95cm}
\subfigure{
  \includegraphics[scale=0.16]{torque_taun0p145_gdot1p5.eps}
\label{fig:test:a}
}
\hspace{0.2cm}
\subfigure{
\includegraphics[scale=0.27]{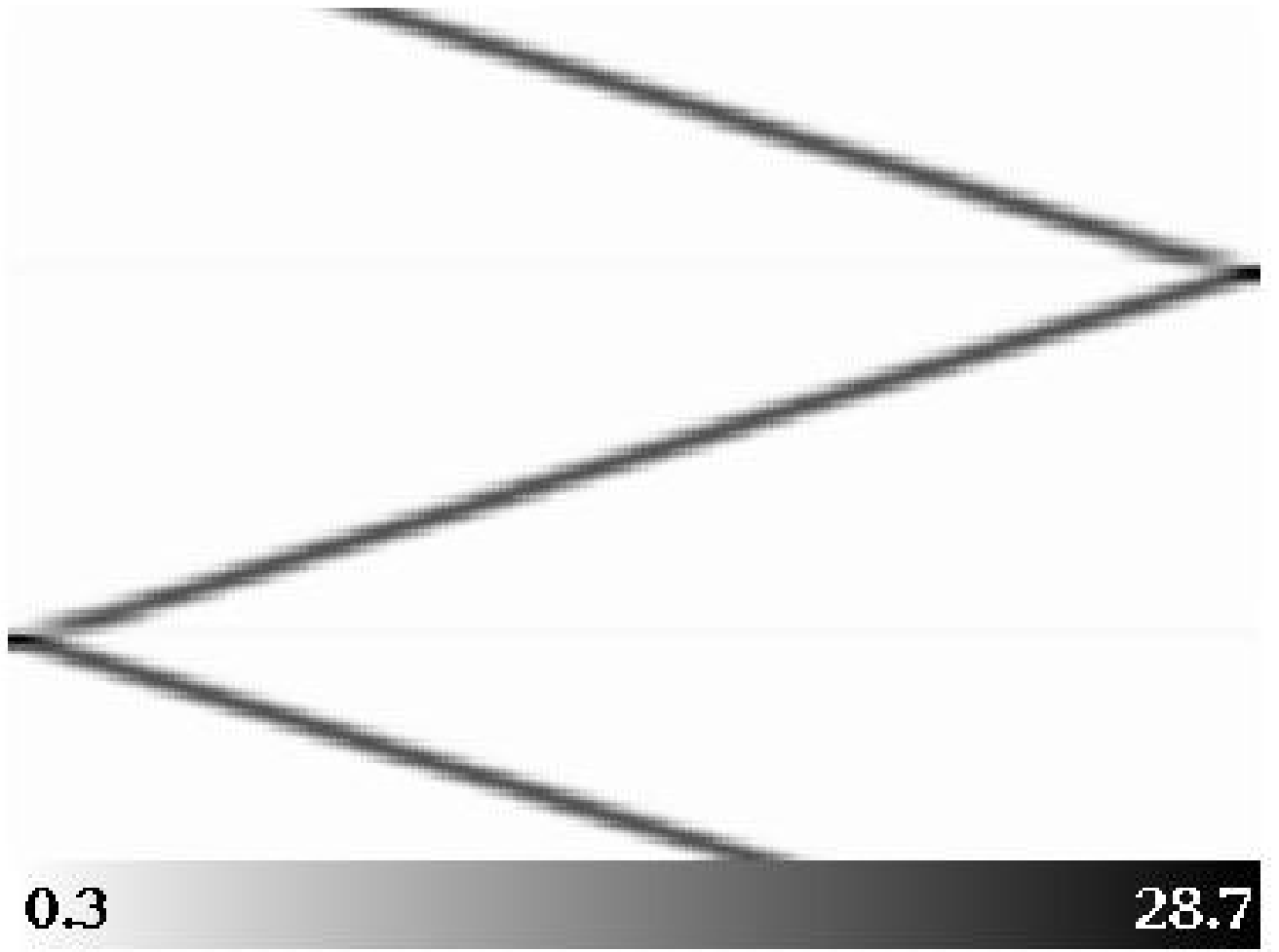}
\label{fig:test:b}
}\\
\vspace{-0.6cm}
\hspace{-1.15cm}
\subfigure{
  \includegraphics[scale=0.16]{torque_taun0p145_gdot7p0.eps}
\label{fig:test:c}
}
\hspace{0.2cm}
\subfigure{
  \includegraphics[scale=0.245]{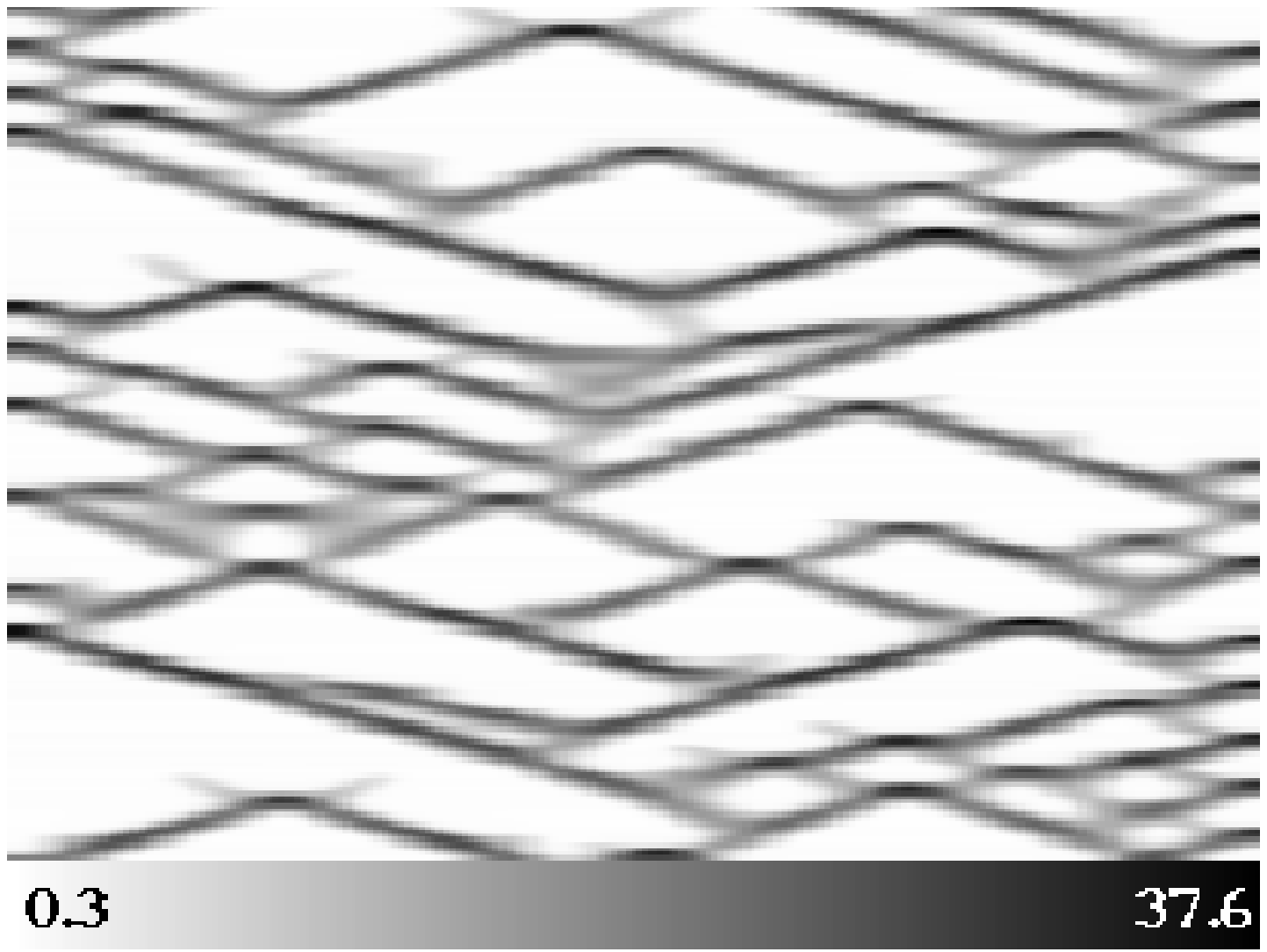}
\label{fig:test:d}
}\\
\vspace{-0.6cm}
\hspace{-1.15cm}
\subfigure{
\includegraphics[scale=0.16]{torque_taun0p145_gdot11p35.eps}
\label{fig:test:e}
}
\hspace{0.2cm}
\subfigure{
  \includegraphics[scale=0.27]{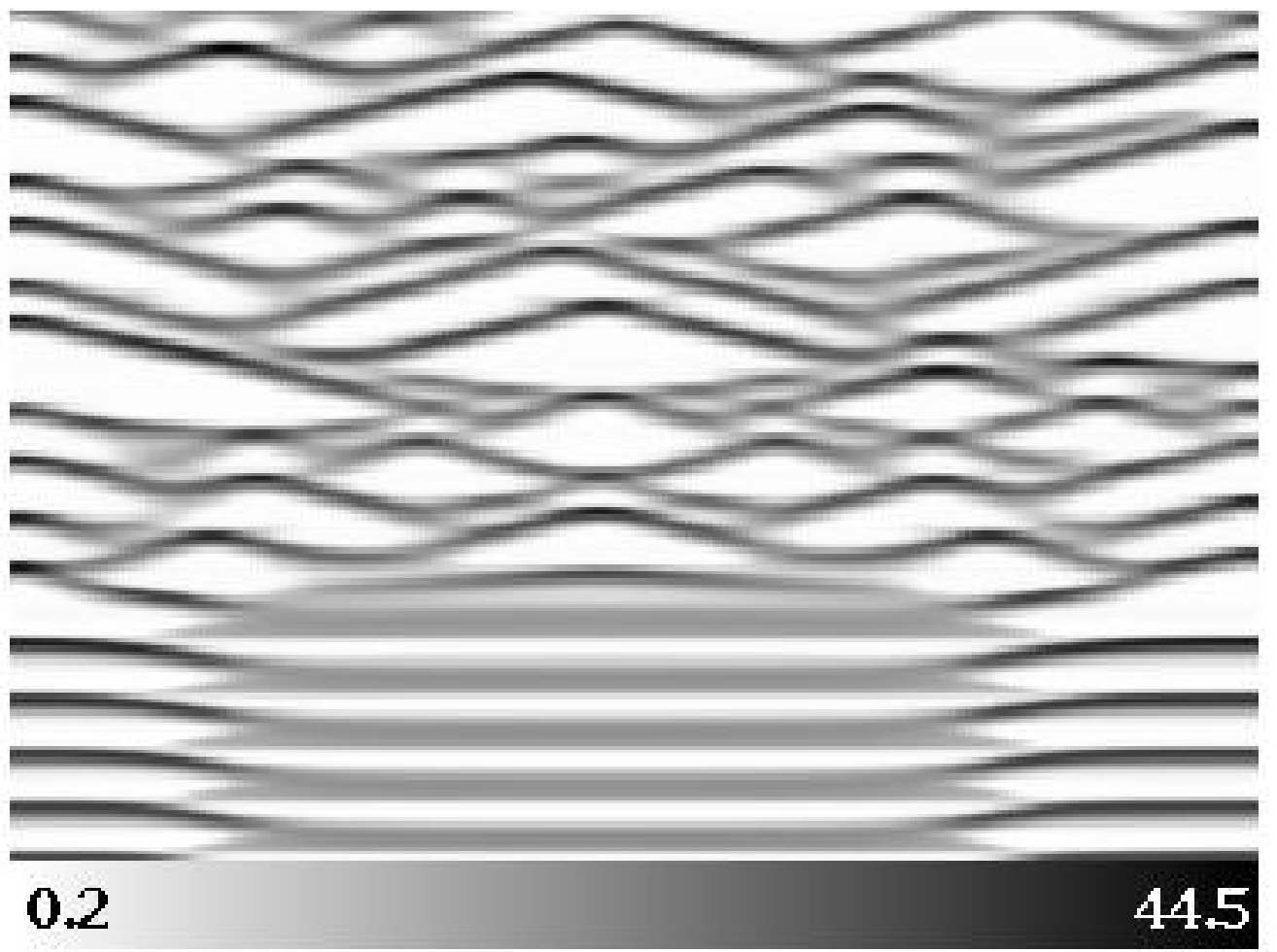}
\label{fig:test:f}
}\\
\vspace{-0.6cm}
\hspace{-1.15cm}
\subfigure{
  \includegraphics[scale=0.16]{torque_taun0p145_gdot19p2.eps}
\label{fig:test:g}
}
\hspace{0.2cm}
\subfigure{
\includegraphics[scale=0.27]{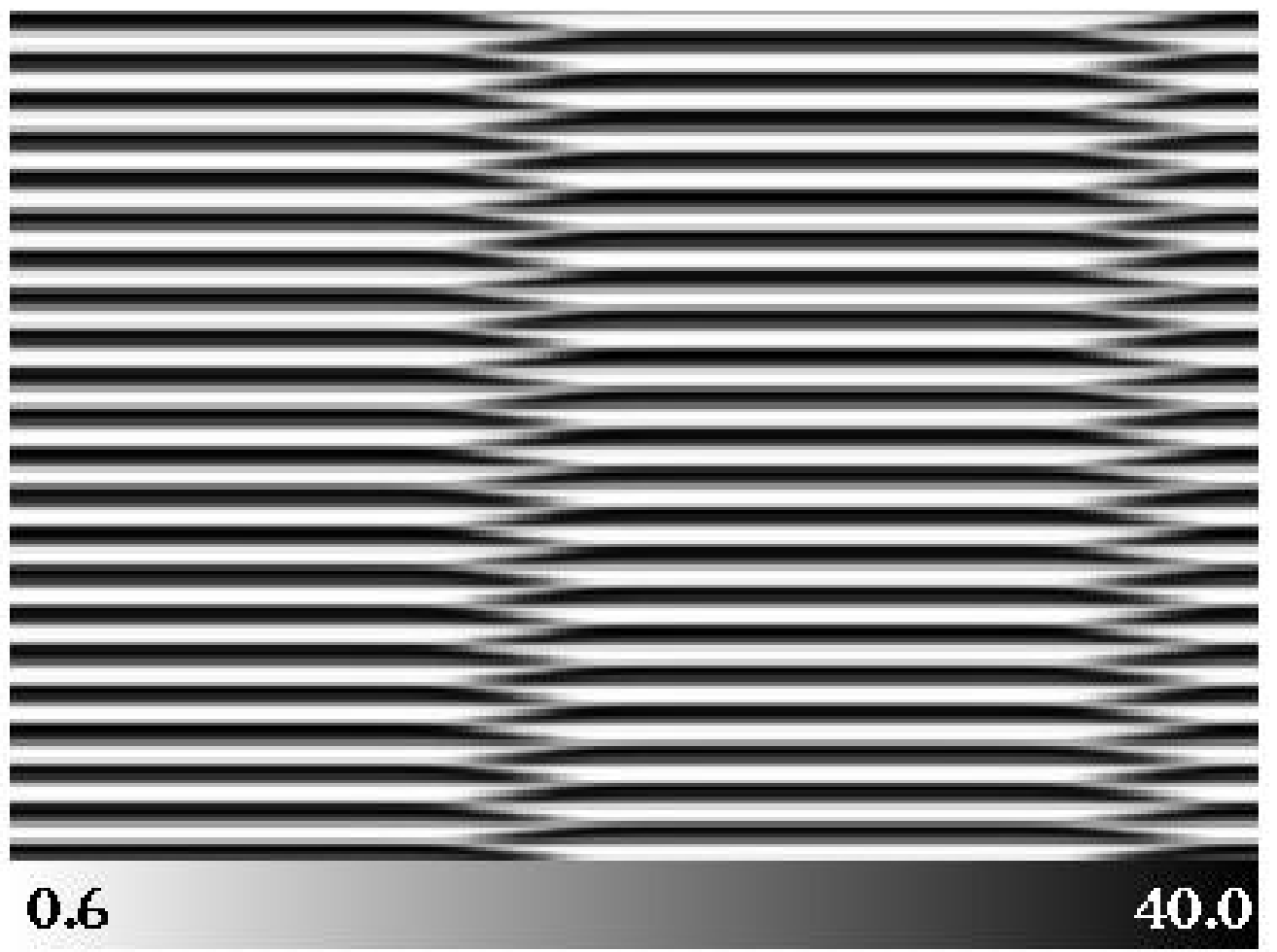}
\label{fig:test:h}
}
\\
\vspace{-0.6cm}
\hspace{-1.15cm}
\subfigure{
  \includegraphics[scale=0.16]{torque_taun0p145_gdot23p0.eps}
\label{fig:test:i}
}
\hspace{0.2cm}
\subfigure{
  \includegraphics[scale=0.245]{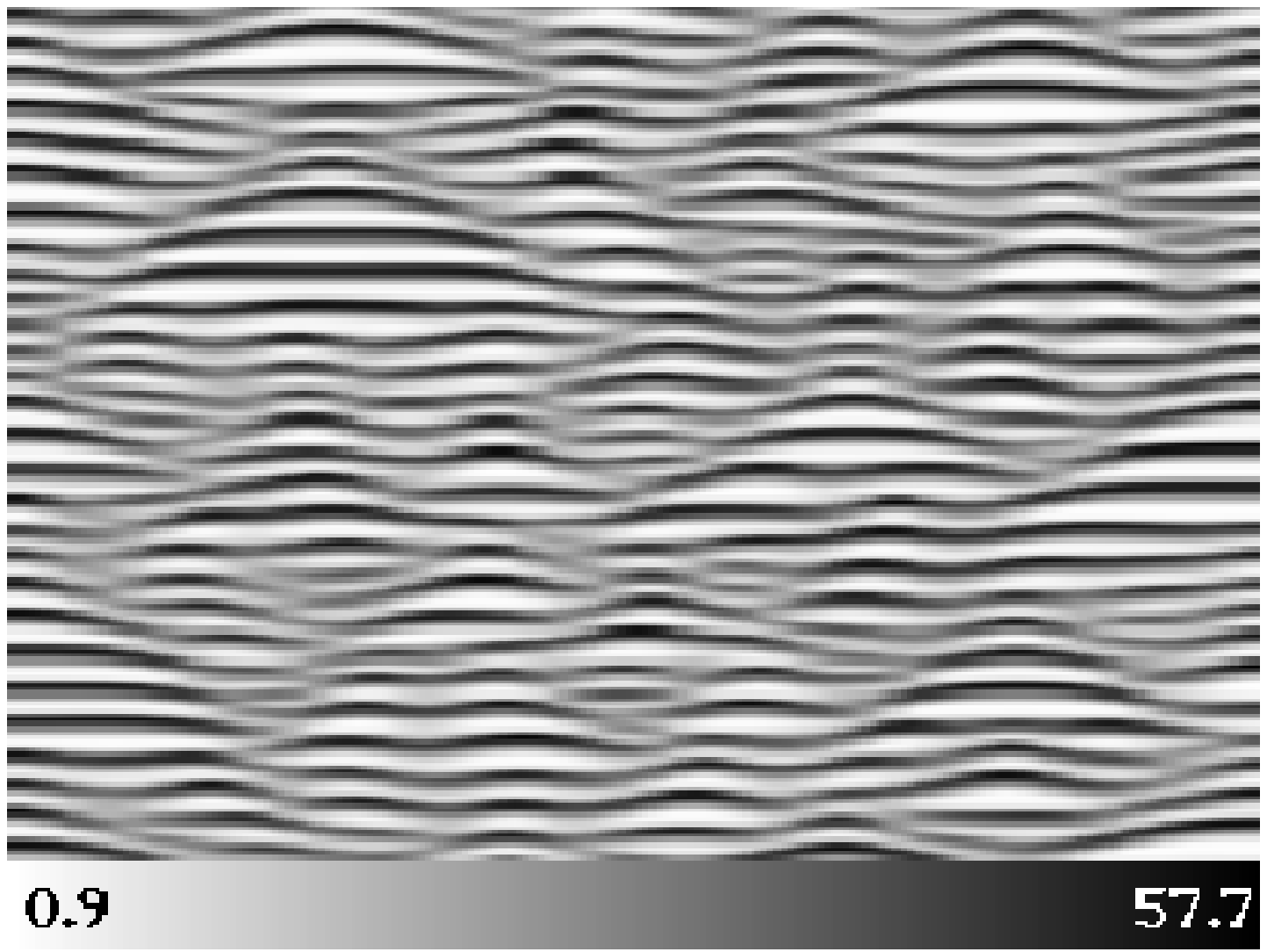}
\label{fig:test:j}
}\\
\vspace{-0.6cm}
\hspace{-1.15cm}
\subfigure{
\includegraphics[scale=0.16]{torque_taun0p145_gdot31p0.eps}
\label{fig:test:k}
}
\hspace{0.2cm}
\subfigure{
  \includegraphics[scale=0.27]{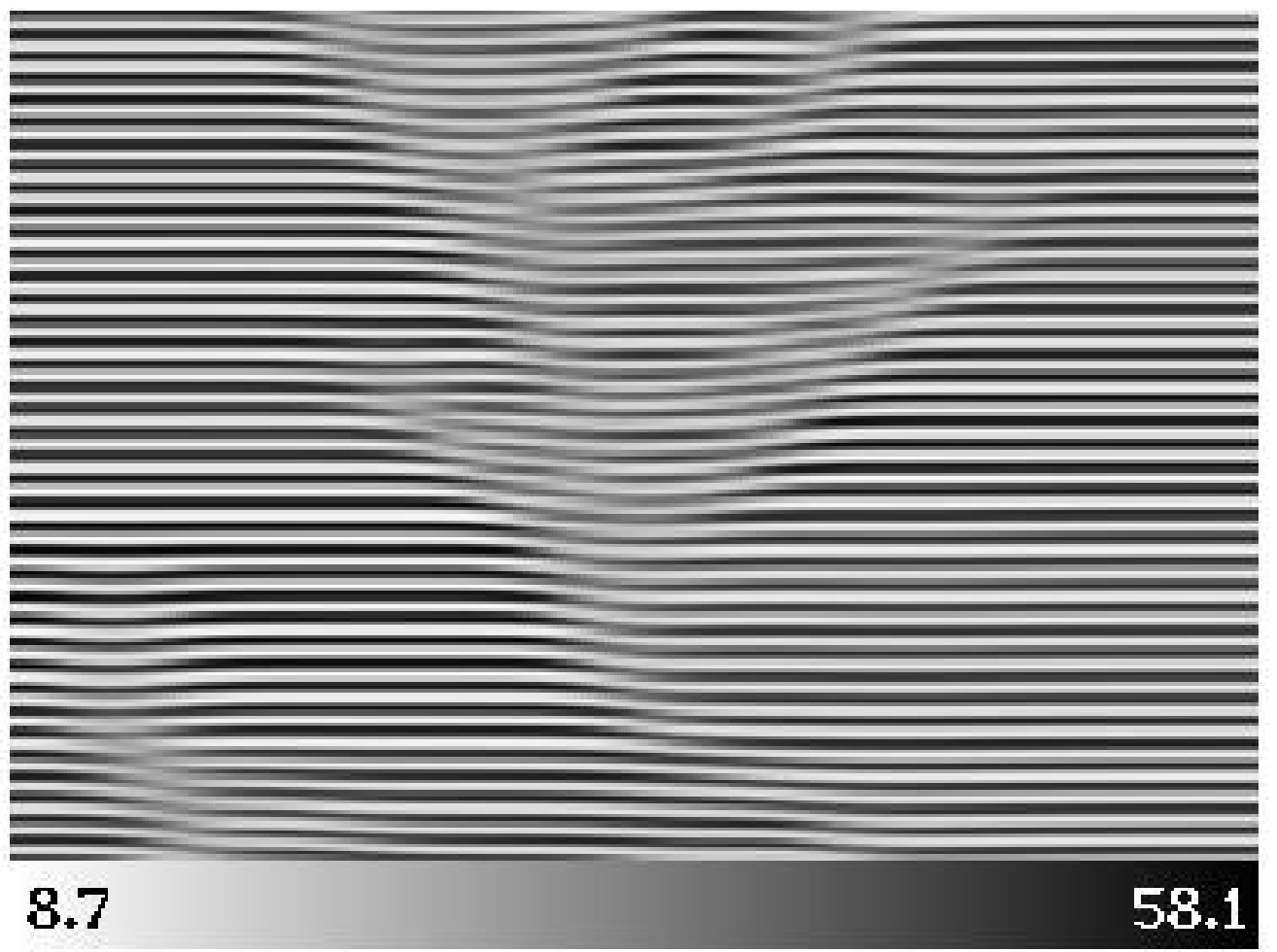}
\label{fig:test:l}
}
\caption{Right: space time plots for shear rate evolution for $\tau_n=0.145$, $D=0.0016$ with space $y=0-1$ left-right and time  $t=120-140$ bottom-top. Greyscale for the shear rate is  shown in each case.  Left: corresponding stress \versus\ time.}
\label{fig:taun0.145}
\vspace{-0.5cm}
\end{figure}

{\it Spatially heterogeneous dynamics---} We now turn to the non-local
model, $D\neq0$, focusing on the implications of an unstable high
shear branch (Fig.~\ref{fig:linear}b,c). The dimensionality $d$ is now
effectively infinite, since each spatial point has its own value of
$n$ and $\gdot$.  We solved the non-local equations using a
Crank-Nicholson algorithm (checking our results with the Rosenbrock
method)~\cite{PreTeuVetFla92}, with the constraint of fixed global
strain rate, $\bar{\dot\gamma}=\int_0^1dy\,\dot{\gamma}(y,t)$. For
small $\tau_n\alt 0.115$ we find stable shear bands
(Fig.~\ref{fig:linear}a). In contrast for $\tau_n=0.145$ (unstable
high shear branch) we find spatio-temporal oscillations and
chaotic banding flows (Fig.~\ref{fig:taun0.145}).

Several regimes are evident. At low applied shear, $\gdotbar=1.5$, a
thin pulse of high shear ricochets back and forth across the cell
(Fig.~\ref{fig:taun0.145}, top). A thin fluctuating high shear band,
away from the rheometer wall, was seen experimentally in
Ref.~\cite{callnote1}.  At larger shear rates, we find two or more
such pulses. For two pulses (not shown), we typically find a
periodically repeating movie with the pulses alternately bouncing off
each other (mid-cell) and the cell walls.  Once three pulses are
present, $\gdotbar=7.0$, periodicity gives way to chaotic behaviour
(Fig.~\ref{fig:taun0.145}).

At still higher shear, $\gdotbar=19.2$, we find regular oscillations
of spatially extended bands pinned at a stationary defect. The local
shear rates span both the low and high shear constitutive branches.
Oscillating (vorticity) bands were seen experimentally in
Ref.~\cite{FisWheFul02,HilVla02}. For the intermediate value
$\gdotbar=11.35$ we find intermittency between patterns resembling
those for $\gdotbar=7.0$ and $\gdotbar=19.2$.

\begin{figure}[t]
  \includegraphics[scale=0.28]{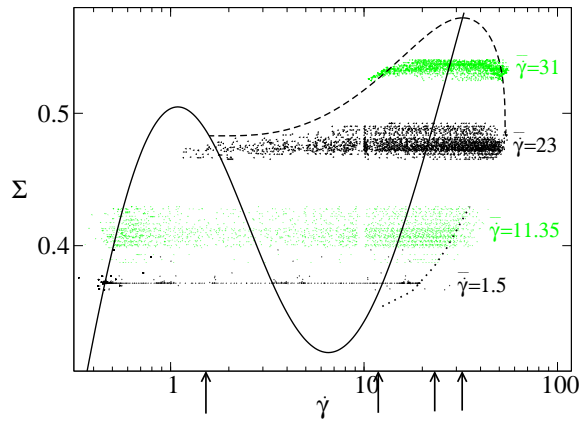}
\caption{Space-time data of Fig.~\ref{fig:taun0.145} for
  $\gdotbar=1.5,11.35,23.0,31.0$ (arrows under abscissa) and
  $130<t<140$ in condensed form.  Solid line -- intrinsic constitutive
  curve; long dashed line -- extrema of the periodic orbits of the
  local model; dotted line -- maximum shear rate of the propagating
  high shear pulse described by Eqns.~\ref{eqn:3d}. 
\label{fig:showall} } 
\end{figure}

Finally for $\gdotbar=23.0,31.0$ we find oscillating bands separated
by {\em moving} defects (Fig.~\ref{fig:taun0.145}, bottom right) with
the flow now governed only by the high shear constitutive branch. In
each band, the shear rate cycles round the periodic orbit of the local
model  (Fig.~\ref{fig:localFlows} and top data set of
Fig.~\ref{fig:showall}).

\begin{figure}[t]
\vspace{-0.5cm}
  \centering 
\hspace{-0.95cm} \subfigure{
    \includegraphics[scale=0.27]{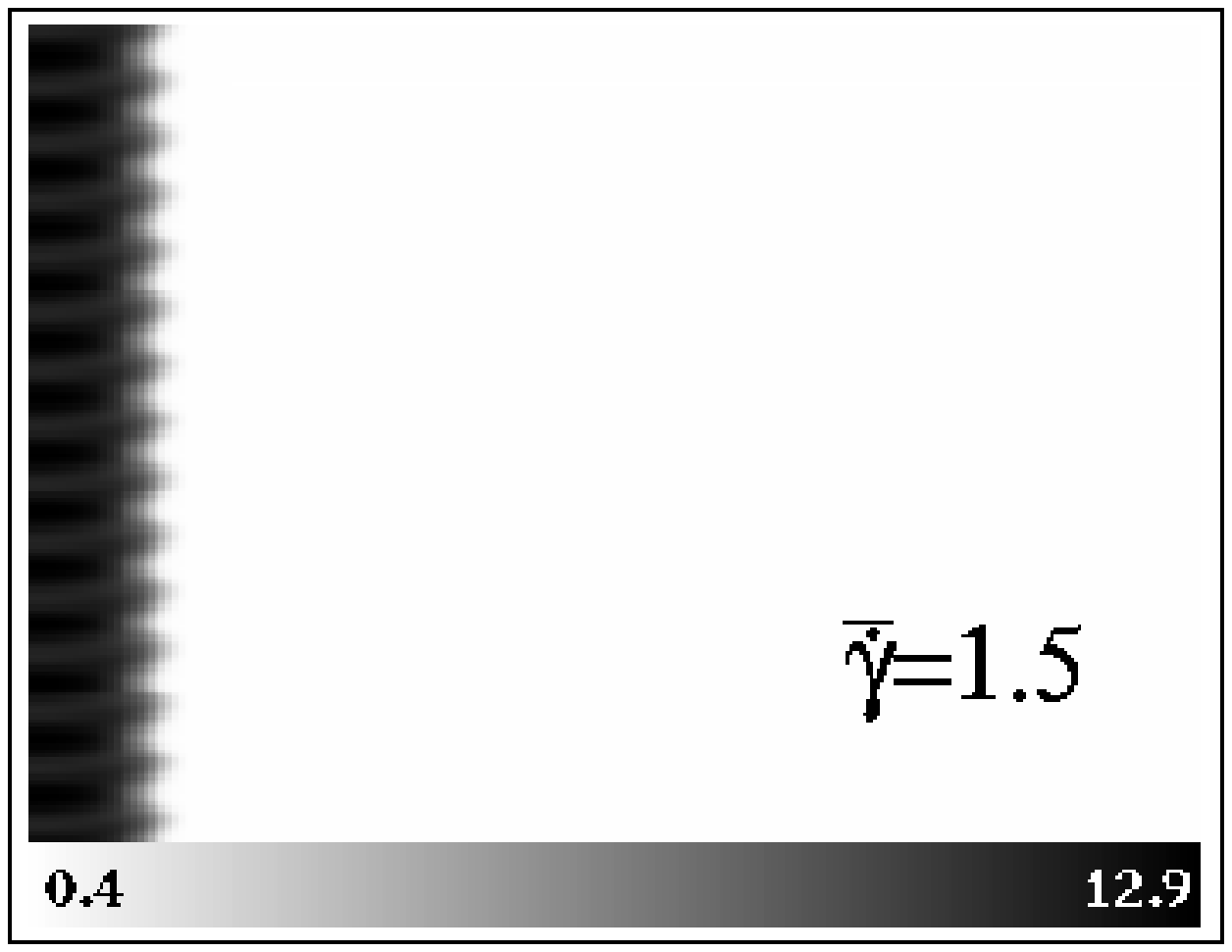}
\label{fig:testing:a}
}
\hspace{0.1cm}
\subfigure{
\includegraphics[scale=0.27]{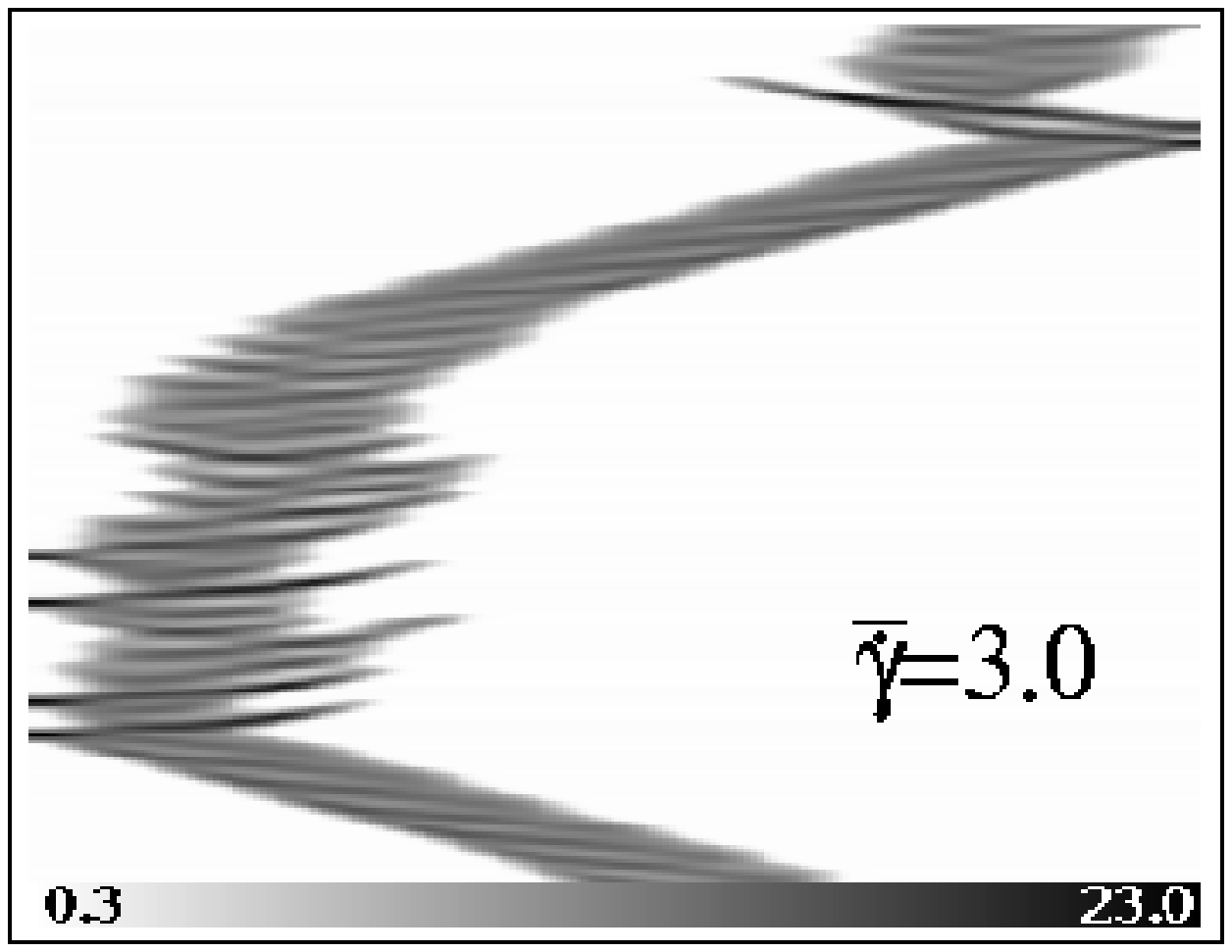}
\label{fig:testing:b}
}\\
\vspace{-0.7cm}
\caption{Shear rate density plots for $\tau_n=0.13$, $D=0.0016$.}
\label{fig:taun0.13}
\vspace{-0.5cm}
\end{figure}

For different values of $\tau_n$ we find a host of other interesting
phenomena~\cite{long}. For example, for weaker instability
($\tau_n=0.13$) at low applied shear rates we see a high shear band
that can pulsate in width while adhering to the rheometer wall
(Fig.~\ref{fig:taun0.13}a), or meander about the cell
(Fig.~\ref{fig:taun0.13}b).  The former behaviour resembles
interfacial motion in seen WMs~\cite{callnote2,HBP98} and onion
phases~\cite{WunColLenArnRou01}.

Finally we discuss in more detail the high shear pulse of
Fig.~\ref{fig:taun0.145} (top right).  At times when the pulse is far
from the wall, the stress $\Sigma$ is constant (flat regions in
Fig.~\ref{fig:taun0.145}, top left). In this regime we transform to the
pulse's comoving frame $\hat{y}=y-ct$ and eliminate $\sigma$ to get
\bea 
c\eta\gdot'&=&-\frac{\Sigma}{\tau(n)}+\frac{g[\gdot\tau(n)]+\eta\gdot}{\tau(n)}-D\eta\gdot''\\
-cn'&=&-\frac{n}{\tau_n}+\frac{N(\gdot\tau_n)}{\tau_n}
\label{eqn:3d}
\eea 
in the $3d$ space $(n,\gdot,\gdot')$, with parameters $\Sigma$ and
$c$.  (Prime denotes $\partial/\partial \hat{y}$.) The low shear
regions in Fig.~\ref{fig:taun0.145} (top right) lie on the low shear
branch of the constitutive curve. Although this is a stable stationary
flow branch (Fig.~\ref{fig:linear}), in the space of
$(n,\gdot,\gdot')$ it is actually a fixed point ($P$) with one
unstable eigenvector and two stable ones. As $\hat{y}$ increases
through the high shear pulse, the solution's trajectory moves away
from $P$ along the unstable direction and then back to $P$ in the
stable plane, completing a ``homoclinic
connection''~\cite{Strogatz94}.  The requirement for this $1d$
trajectory to correctly rejoin the $2d$ stable manifold selects (for a
given $\Sigma$) the observed propagation speed, $c$. For other values
of $c$, the trajectory abruptly crosses the $2d$ manifold, missing
$P$. (This follows by simple dimension counting~\cite{lu99}.)  Once
selected, the solution's profile $\gdot(\hat{y})$ can (for a given
value of $D$) be integrated over $\hat{y}$ to find the globally
applied shear rate $\gdotbar$.  By the reverse reasoning, we find a
selected $\Sigma$ and $c$ for any $\gdotbar$ (and $D$).  We used
AUTO~\cite{AUTO} to check that the homoclinic connections of
Eqn.~\ref{eqn:3d} indeed coincide with our numerical results.  The
locus of the maxima $\gdot_{\rm h}(\Sigma)$ of these connections is
marked in Fig.~\ref{fig:showall}.  The space-time points for
$\gdot>\gdot_{\rm h}$ arise during wall collisions; correspondingly,
their density is smaller than for $\gdot<\gdot_{\rm h}$.

To summarise, we have constructed a simple model of shear banding in
which the high shear branch of the underlying constitutive curve is
rendered unstable by a coupling between flow and microstructure.
Within this model, we have found a rich variety of spatio-temporal
oscillatory and rheochaotic flows, many resembling experimental
observations in shear banding systems. It remains an open challenge to
delineate more fully the spectrum of mechanisms governing rheochaotic
banded states.  Extension to higher dimensions, allowing fluctuations
along the plane of the interface would be interesting.  Within the
present one-dimensional approach, generalisation to shear thickening
systems~\cite{catesnote}, curved Couette geometries and
vorticity-banding is underway~\cite{long}.

We thank A. A. Aradian, P. T.  Callaghan, M. E. Cates, E. Knobloch, A.
M.  Rucklidge and J.-B. Salmon for interesting discussions; and EPSRC
GR/N11735 for funding.

\vspace{-0.3cm}


\end{document}